\documentclass[12pt]{article}
\usepackage{amsmath}

\begin{document}
\title{Exact solution of generalized Dirac oscillator in electric field}

\author{
H. P. Laba$^1$,
V. M. Tkachuk$^2$\footnote{voltkachuk@gmail.com}\\
$^1$Department of Applied Physics and Nanomaterials Science, \\
Lviv Polytechnic National University,\\
5 Ustiyanovych St., 79013 Lviv, Ukraine,\\
$^2$Department for Theoretical Physics,\\
Ivan Franko National University of Lviv,\\
12, Drahomanov St., Lviv, 79005, Ukraine.}

\maketitle

\begin{abstract}
We study generalized (1+1)-dimensional Dirac oscillator in nonuniform electric field.
It is shown that in the case of specially chosen electric field
the eigenvalue equation can be casted in the form of supersymmetric quantum mechanics.
It gives a possibility to find exact solution for the energy spectrum of the generalized Dirac oscillator in nonuniform electric field.
Explicit examples of exact solutions  are presented. We show that sufficiently large electric field destroys the bounded eigenstates.

Key words: generalized Dirac oscillator, supersymmetric quantum mechanics, exact solution.

PACS numbers: 03.65.-w,  03.65.Pm.
\end{abstract}

\section{Introduction}

The Dirac oscillator is one of  few relativistic
systems admitting exact solutions. For the first time the oscillator
was introduced in \cite{Ito67}
as relativistic system with Hamiltonian linear in both momenta and coordinates.
Later this relativistic system was reintroduced in \cite{Mosh89} where it was called as the Dirac oscillator.
This publication immediately gave rise to much interest in studies of the system. It is worth mentioning that experimental realization of the Dirac oscillator was proposed recently in \cite{Fra13}. Short history of  Dirac oscillator can be found in \cite{Que17}.

Note that the Dirac oscillator can be solved  exactly in the case
of quantized space with minimal length \cite{Que05,Que06}. There is also generalized version of
Dirac oscillator which is exactly solvable \cite{Hoa04,Ikh12,Dut13,Jun17}.

In this paper we propose new class of exactly solvable relativistic systems.
Namely, we show that the generalized Dirac oscillator can be  exactly solved in the case of specially
chosen nonuniform electric field. The paper is organized as follows. In Section 2 we consider the generalized Dirac oscillator in nonuniform electric field and find its relation with supersymmetric (SUSY) quantum mechanics. This gives a possibility to find exact solution of eigenvalue equation.
In Section 3 we present examples of generalized Dirac oscillators in electric field which can be solved exactly.

\section{Generalized Dirac oscillator in electric field and SUSY quantum mechanics}
Eigenvalue equation for (1+1)-dimensional Dirac oscillator
reads
\begin{eqnarray}
(\sigma_x(p_x-im\omega x\sigma_z)+m\sigma_z)\psi=E\psi
\end{eqnarray}
where we put the Planck constant $\hbar=1$ and the velocity of light  $c=1$. The momentum operator reads $p_x=-id/dx$.

Generalized Dirac oscillator can be obtained from the ordinary Dirac oscillator with the help of substitution
\begin{eqnarray}
m\omega x\to W(x),
\end{eqnarray}
where $W(x)$ can be called as superpotential.

In addition we take into consideration the
potential energy  $U(x)=eV(x)$ of charge $e$ in scalar potential $V(x)$ which corresponds to nonuniform electric field $-dV(x)/dx$.
Eigenvalue equation for generalized Dirac oscillator in scalar potential now reads
\begin{eqnarray}
(\sigma_x(p_x-iW(x)\sigma_z)+m\sigma_z)\psi=(E-U(x))\psi
\end{eqnarray}
which can be rewritten in the form
\begin{eqnarray}\label{GDO}
(\sigma_x p_x-\sigma_y W(x)+m\sigma_z-(E-U(x))\psi=0.
\end{eqnarray}

In order to reduce this equation to second order one we present solution in the form
\begin{eqnarray}
\psi=(\sigma_x p_x-\sigma_y W(x)+m\sigma_z+(E-U(x))\tilde\psi.
\end{eqnarray}
 Substituting it into (\ref{GDO}), we find equation for new function $\tilde\psi$
\begin{eqnarray}\label{DEtilde}
(p_x^2+W^2+m^2-(E-U)^2-\sigma_zW'+i\sigma_x U')\tilde\psi=0,
\end{eqnarray}
where prime denotes the derivative $W'(x)=dW(x)/dx$.

Let us consider special case
\begin{eqnarray}\label{UW}
U=\kappa W,
\end{eqnarray}
where $\kappa$ is some real constant.
In this case equation (\ref{DEtilde}) reads
\begin{eqnarray}\label{DEtilde1}
(p_x^2+W^2+m^2-(E-\kappa W)^2+(-\sigma_z+i\kappa\sigma_x)W')\tilde\psi=0.
\end{eqnarray}
It is important that the spin operators are included now in one block $(-\sigma_z+i\kappa\sigma_x)$ and solution can be found in the form
\begin{eqnarray}
\tilde\psi=\chi\phi(x),
\end{eqnarray}
where $\chi$ is spin part of wave function satisfying equation
\begin{eqnarray}
(-\sigma_z+i\kappa\sigma_x)\chi=\lambda \chi
\end{eqnarray}
with eigenvalues
\begin{eqnarray}
\lambda=\sigma \sqrt{1-\kappa^2},
\end{eqnarray}
here $\sigma=\pm 1$.

Equation for coordinate part of the wave function can be written in the form
\begin{eqnarray}
(p_x^2+(1-\kappa^2)W^2+2E\kappa W+ \sigma \sqrt{1-\kappa^2} W')\phi=(E^2-m^2)\phi.
\end{eqnarray}

This equation can be reduced to the form of equation of  SUSY quantum mechanics
\begin{eqnarray}\label{SUSY}
(p_x^2+\tilde W^2+\sigma \tilde W')\phi=\epsilon\phi.
\end{eqnarray}
where
\begin{eqnarray}\label{epsilon}
\epsilon=\left({E^2\over 1-\kappa^2}-m^2\right)
\end{eqnarray}
and
\begin{eqnarray}\label{tildeW}
\tilde W=\sqrt{1-\kappa^2}\left(W+{\kappa\over 1-\kappa^2}E\right)
\end{eqnarray}
 can be treated as  new superpotential. For review of SUSY quantum mechanics see for instance \cite{Coo95,Gan11}.

\section{Exact solution of Dirac oscillators for different electric fields}

Equation (\ref{SUSY}) can be solved exactly for the so called shape-invariant superpotentials \cite{Gen83}.
Note that all exactly solvable potentials in one-dimensional quantum mechanics correspond to this case.

{\it Example 1}.
Let us consider $W=x$ that leads to the ordinary Dirac oscillator with $\omega m=1$.
In this case the potential energy $U=\kappa W=\kappa x$
gives the uniform electric field, strength of which multiplied by charge $e$ is $-\kappa$.
Equation (\ref{SUSY}) can be written in the form of eigenvalue equation for harmonic oscillator
\begin{eqnarray}\label{SUSYho}
(p_x^2+(1-\kappa^2)(x-x_0)^2+\sigma\sqrt{1-\kappa^2})\phi=\epsilon\phi,
\end{eqnarray}
where $x_0={\kappa E/\sqrt{1-\kappa^2}}$.
Energy spectrum of this harmonic oscillator is well know
\begin{eqnarray}
\epsilon_{\sigma}=\sqrt{(1-\kappa^2)}\left(2n+1+\sigma\right), \ \ n=0,1,2,...
\end{eqnarray}
Then according to (\ref{epsilon}) energy spectrum of Dirac oscillator in this case is as follows
\begin{eqnarray}
E_{\sigma}^{\pm}=\pm\sqrt{(1-\kappa^2)(\sqrt{1-\kappa^2}(2n+1+\sigma)+m^2)}.
\end{eqnarray}
Introducing $n_{\sigma}=n+(1+\sigma)/2$, the energy spectrum can be written in the form
\begin{eqnarray}
E_{\sigma}^{\pm}=\pm\sqrt{(1-\kappa^2)(\sqrt{1-\kappa^2}2n_{\sigma}+m^2)},
\end{eqnarray}
where $n_-=0,1,2,...$, $n_+=1,2,...$.
One can see that nonzero energy levels are two-fold degenerated when $n_+=n_-=1,2,...$. This is result of
supersymmetry.
It is interesting to note that
when the value of electric field is large then
critical one $|\kappa|>1$ the bounded eigenstates
are absent.

{\it Example 2}. Let us consider
$W=\alpha_0\tan(x)$, where $\alpha_0>0$.
Then potential energy (\ref{UW}) reads $U(x)=\kappa\alpha_0\tan(x)$ and corresponds to the nonuniform electric field.
In this case we have
\begin{eqnarray}\label{tan}
\tilde W= \alpha\tan(x)+\beta,
\end{eqnarray}
where
\begin{eqnarray}\label{ea}
\alpha=\alpha_0\sqrt{1-\kappa^2},\ \ \beta={\kappa\over\sqrt{1-\kappa^2}}E.
\end{eqnarray}
Eigenvalue of (\ref{SUSY}) with superpotential (\ref{tan}) is well known \cite{Coo95,Gen83}
\begin{eqnarray}
\epsilon_{\sigma}=\alpha^2+\beta^2-(\alpha-n_{\sigma})^2-{\alpha^2\beta^2\over(\alpha-n_{\sigma})^2},
\end{eqnarray}
where $n_{-}=0,1,2,...,n_{\rm max}<\alpha$ for $\sigma=+1$ and $n_{+}=1,2,...,n_{\rm max}<\alpha$ for $\sigma=-1$.
Then taking into account (\ref{epsilon}) and (\ref{ea}) we obtain the following equation for the energy spectrum
\begin{eqnarray} \nonumber
E_{\sigma}^2+{\alpha_0^2\kappa^2\over (\alpha_0\sqrt{1-\kappa^2}-n_{\sigma})^2}E^2_{\sigma}=\\
=m^2+\alpha_0^2(1-\kappa^2)
-(\alpha_0\sqrt{1-\kappa^2}-n_{\sigma})^2
\end{eqnarray}
solution of which give the energy spectrum for generalized Dirac oscillator
\begin{eqnarray}
E^{\pm}_{\sigma}=
\pm
\sqrt{{m^2+\alpha_0^2(1-\kappa^2)
-(\alpha_0\sqrt{1-\kappa^2}-n_{\sigma})^2\over 1+\alpha^2_0\kappa^2/(\alpha_0\sqrt{1-\kappa^2}-n_{\sigma})^4}}.
\end{eqnarray}
As result of supersymmetry non-zero energy levels are two-fold degenerated at $n_-=n_+=1,2,...$.
Concerning relation of degeneracy of energy levels with algebra of supersymmetry see for instance papers \cite{Coo95,Tka97}.

Similarly as in the first example at  $|\kappa|>1$ the bounded eigenstates are absent.
So, the sufficiently strong electric field destroys the bounded states.

\section{Conclusion}
In this paper we proposed new class of exactly solvable relativistic systems.
We found that generalized (1+1)-dimensional Dirac oscillator in electric field can be solved exactly when superpotential $W(x)$
and potential energy are related linearly as in (\ref{UW}). We showed that in this case the  eigenvalue equation
for considered system can be reduced to the eigenvalue equation of SUSY quantum mechanics with new superpotential
$\tilde W(x)$ given by (\ref{tildeW}). For shape invariant superpotentials the generalized Dirac oscillator in electric field is exactly solvable.
In contrast to the ordinary SUSY the superpotential $\tilde W(x)$ contains energy of the system. Therefore applying technique of the shape-invariant potentials   we obtain some algebraic
equation the solution of which gives the exact energy spectrum.
In addition the superpotential $\tilde W(x)$ becames imaginary for $|\kappa|>0$ that corresponds to
large electric field. In this case the bound eigenstates are absent. Thus, sufficiently strong electric field destroy bound eigenstates which exist only for $|\kappa|<0$ when superpotential is real.
Note also that as a result of SUSY
all nonzero energy levels are two-fold degenerated.

We presented examples of exact solutions of the problem.
In the first example we considered ordinary Dirac oscillator in uniform electric field.
 The energy spectrum for this case was found. In contrast to the  non-relativistic harmonic oscillator the energy spectrum for this example essentially depends on the strength of electric field. New feature in this case is that sufficiently strong electric field destroys the bounded states. In relation with this it is worth to mention the paper \cite{Nat14}
where
the authors study (2 + 1) dimensional massless Dirac oscillator in the
presence of perpendicular magnetic and transverse electric fields.
The authors found  essential dependence of spectrum on the magnetic field, namely,
there exists a
critical magnetic field such that the spectrum is different in the
two regions for magnetic field that less critical value and magnetic field that large critical value.
As a second example we considered the generalized Dirac oscillator in non-uniform electric field. In this case the sufficiently strong electric field also destroys the bounded states. So, choosing different shape invariant superpotentials one obtain new exactly solvable generalized Dirac oscillators in the electric fields.

\section*{Acknowledgments}
This work was supported in part  by the project $\Phi\Phi$-30$\Phi$ (No. 0116U001539) from the Ministry of Education and Science of Ukraine.

\end{document}